\documentclass[12pt]{iopart}
\usepackage{iopams,graphicx,texdraw,epsf,bm}

\newcommand{\be}{\begin{equation}}
\newcommand{\ee}{\end{equation}}
\newcommand{\bea}{\begin{eqnarray}}
\newcommand{\eea}{\end{eqnarray}}

\begin{document}
\paper[Compatibility of $1/n$ and $\epsilon$ expansions\ldots]
{Compatibility of $\boldsymbol{1/n}$ and
$\boldsymbol{\epsilon}$ expansions for critical exponents at
$\boldsymbol{m}$-axial Lifshitz points}
\author[M A Shpot \etal ]{M A Shpot{\dag}{\ddag}, H W Diehl{\ddag}
  and Yu M Pis'mak{\S}{\ddag}}
\address{{\dag}\ Institute for Condensed Matter Physics, 79011 Lviv, Ukraine}
\address{{\ddag}\ Fachbereich Physik, Universit{\"a}t Duisburg-Essen,
D-47048 Duisburg, Germany}
\address{{\S}\ State University of Sankt-Petersburg, 198504
  Sankt-Petersburg, Russia}
\ead{shpot@ph.icmp.lviv.ua}
\setcounter{footnote}{3}
\begin{abstract}
  The critical behaviour of $d$-dimensional $n$-vector models at
  $m$-axial Lifshitz points is considered for general values of $m$ in
  the large-$n$ limit. It is proven that the recently obtained
  large-$n$ expansions [J.\ Phys.: Condens.\ Matter {\bf 17}, S1947
  (2005)] of the correlation exponents $\eta_{L2}$, $\eta_{L4}$ and
  the related anisotropy exponent $\theta$ are fully consistent with
  the dimensionality expansions to second order in $\epsilon=4+m/2-d$
  [Phys.\ Rev.\ B {\bf 62},  12338  (2000); Nucl.
  Phys. B {\bf 612}, 340 (2001)] inasmuch as both expansions yield
  the same contributions of order $\epsilon^2/n$.
\end{abstract}
\pacs{05.20.-y, 11.10.Kk, 64.60.Ak, 64.60.Fr}
\submitto{\JPA}

\section{Introduction}

Lifshitz points (LP) are familiar examples of multi-critical points.
At a LP a disordered, a homogenous ordered, and a modulated ordered
phases meet \cite{HLS75a,Hor80,Sel92,Die02}. In the case of systems
with $m$-axial LP, there is a degeneracy such that long-range order
modulated along any of $m$ distinct axes can occur in the modulated ordered
phases. The study of critical behavior at such LP began immediately
after their discovery in the middle of the 1970s.

Unfortunately, the technical difficulties one is faced with in
analytical renormalization group (RG) calculations are enormous.  This
is the main reason why RG results based on systematic expansions, such
as expansions in powers of $\epsilon = d^*(m)-d$ about the upper
critical dimension $d^*(m)=4+m/2$, or in powers of $1/n$, where $n$ is
the number of components of the order parameter, had remained quite
scarce for decades. Furthermore, early $\epsilon$-expansion results
obtained by two different groups (cf.\ \cite{SG78} and
\cite{Muk77,HB78}) had yielded contradictory results, and these
discrepancies had remained unclarified for many years. The results of
reference \cite{SG78}, which were restricted to the special cases of
bi- and hexa-axial LP $m=2$ and $m=6$, were reproduced twenty
years later by field-theoretic means \cite{MC99}.  However, a full
two-loop RG analysis in $d^*(m)-\epsilon$ dimensions was reported only
in 2001 \cite{DS00a,SD01}. This gave the $\epsilon$ expansions to
second order of all four main independent critical exponents as well as the
correction-to-scaling exponent for general values of $m$, besides
resolving the mentioned discrepancies.\footnote{Alternative results
reported in reference \cite{AL01} could be refuted \cite{DS01a}.}

In a recent paper \cite{SPD05} (hereafter referred to as I), we have
shown how the $1/n$ expansion can be applied to the study of critical
behavior at $m$-axial LP. We determined the correlation exponents
$\eta_{L2}$ and $\eta_{L4}$, and the related anisotropy $\theta$ of
$d$-dimensional systems to first order in $1/n$ for general values of
$m$. The results took the form of complicated integrals whose
integrands involve further multi-dimensional integrals.  We were able to
check that in the isotropic limits $m\to 0$ and $m\to d$ they
correctly reduce to known results, namely, the expansions to order
$1/n$ of the Fisher exponent $\eta$ at a usual critical point
\cite{Ma73,AH73}, and that of $\eta_{L4}$ at the isotropic $d$-axial
LP \cite{HLS75b}, respectively.  Furthermore, we could show that
our $O(1/n)$ results are in conformity with published dimensionality
expansion results about the lower critical dimension $d_*(m)=2+m/2$
both for $m=0$ \cite{BZ76} and for $m=2$ and $m=6$ \cite{GS78}.

The virtues of the $1/n$ expansion are well known: It can be applied
in arbitrary fixed dimensions $d$, does not rely on the smallness of a
further expansion parameter such as $\epsilon$, and yields nontrivial
results below the upper critical dimension $d^*(m)$ in a
mathematically controlled fashion. Besides its capability of providing
valuable information about the critical behavior for given $d$ and
$m$, it allows for nontrivial checks on $\epsilon$-expansion results.
Unfortunately, the complicated general form of our results in
I prevented us from proving their consistency
with the $\epsilon$-expansion results of \cite{DS00a,SD01} for general
$m$. We could verify it in the isotropic limits $m\to 0$ and $m\to d$.
However, the only other case in which we could explicitly demonstrate
this consistency by analytical means was that of $m=2$.  This is
unfortunate for at least two reasons: first, it excludes, in
particular, the physically important uniaxial case $m=1$; second,
unlike both the contributions of first order in $\epsilon=d^*(m)-d$ as
well as their $\Or(d-d_*(m))$ analogues in the expansions about the
lower critical dimensions $d_*(m)$, the $\Or(\epsilon^2)$ terms
exhibit a nontrivial $m$-dependence \cite{DS00a,SD01}. This
propagates into the $\Or(1/n)$ series coefficients, and ought to be
checked.

The purpose of this article is to fill this gap and prove that our
large-$n$ expansion results \cite{SPD05} to order $1/n$ for the
exponents $\eta_{L2}(d,m,n)$, $\eta_{L4}(d,m,n)$, and $\theta(d,m,n)$
are fully consistent with the $\epsilon$-expansion results of
references \cite{DS00a,SD01}. In the next section we first provide the
necessary background, recalling the continuum model on which our
analysis is based as well as our results for general $m$ given in I.
We then show that these results can be rewritten in a form allowing
analytic comparisons with the $\Or(\epsilon^2)$ results of references
\cite{DS00a,SD01} for general $m$. The proof that they are in
conformity with the latter is given in section~\ref{sec:proof}. The
closing section~\ref{sec:concl} contains a brief discussion and
concluding remarks.

\section{Large-$\bm{n}$ expansions of the correlation exponents at the
  Lifshitz point} \label{sec:bgreform}

Just as in I, we consider a model defined by the
Euclidean action
\begin{equation}\label{eq:EH}\fl
{\mathcal H}[\bm{\phi}]
=\frac{1}{2}\,\int\rmd^{d-m}r\int\rmd^mz\left[(\nabla_{\bm{r}}\bm{\phi})^2
+(\nabla_{\bm{z}}^2\bm{\phi})^2
+\tau_{\rm LP}\,\phi^2+\rho_{\rm LP}\,(\nabla_{\bm{z}}\bm{\phi})^2
+\frac{\lambda}{8}\,\phi^4\right].
\end{equation}
Since we intend to work directly at the LP, we have set the
coefficients of the quadratic terms to their corresponding critical
values $\tau_{\rm LP}$ and $\rho_{\rm LP}$ \footnote{In the following
  it is tacitly understood that $d$ with $d<d^*(m)=4+m/2$ and $m$ are
  chosen such that a LP exists. This requires, in particular, that $d$
  exceeds the dimension $2+m/2$ below which the homogeneous ordered
  phase becomes thermally unstable because of spin-wave excitations.
  However, it also requires that the modulated ordered phase remains
  stable. For a discussion of these delicate issues, see the review
  article \cite{Die02} and its references.}.

Here $\bm{\phi}=\bm{\phi}(\bm{x})$ is
the usual $n$-component order-parameter field. Its $d$-dimensional
position vector ${\bm x}=({\bm r},{\bm z})\in \mathbb{R}^d$ has a
$(d-m)$-dimensional ``perpendicular'' component $\bm{r}$ and an
$m$-dimensional ``parallel'' one, $\bm{z}$. The subspace associated
with $\bm{z}$ is the one in which modulated order can occur in the
corresponding phase; that of $\bm{r}$ is its orthogonal complement. A
similar decomposition has been made for the gradient operator
$\nabla=(\nabla_{\bm{r}},\nabla_{\bm{z}})$. Thus $\nabla_{\bm{z}}^2$
is the Laplacian in the parallel subspace.

Employing the notational conventions of I, we
write the wave-vector conjugate to $\bm{x}=(\bm{r},\bm{z})$ as
$\bm{k}=(\bm{p},\bm{q})$, with $\bm{p}\in\mathbb{R}^{d-m}$ and
$\bm{q}\in\mathbb{R}^m$. Further, we introduce the two-point cumulant
$ G_\phi(r,z)$ and its Fourier transform $\tilde{G}_\phi(p,q)$ through
\begin{eqnarray}
  \label{eq:Gphidef}
  G_\phi(r,z)&=&\frac{1}{n}\,\big[\langle\bm{\phi}(\bm{r},\bm{z})
  \cdot\bm{\phi}(\bm{0},\bm{0})\rangle
  -\langle\bm{\phi}(\bm{r},\bm{z})
  \rangle\cdot\langle\bm{\phi}(\bm{0},\bm{0})\rangle\big]\nonumber\\
  &=&\int^{(d-m)}_{\bm{p}}\int^{(m)}_{\bm{q}} \e^{\rmi
    (\bm{r}\cdot\bm{p}+ \bm{z}\cdot\bm{q})}\,\tilde{G}_\phi(p,q)\;,
\end{eqnarray}
where $\int^{(d-m)}_{\bm{p}}\equiv
(2\pi)^{-d+m}\int_{\mathbb{R}^{d-m}} d^{d-m}p$ and $\int_{\bm{q}}^{(m)}\equiv
(2\pi)^{-m}\int_{\mathbb{R}^m} d^mq$ denote normalized  $(d-m)$- and
$m$-dimensional integrals, respectively.

As discussed in I, the full propagator
$\tilde{G}_\phi(p,q)$ becomes a generalized homogeneous function
$\tilde{G}^{({\rm as})}_\phi(p,q)$ in the limit of large
length-scales. The latter function satisfies at the LP the
homogeneity relations
\begin{equation}\label{eq:Gold}
\tilde{G}^{({\rm as})}_\phi(p,q)=
p^{-2+\eta_{L2}}\,\tilde{G}^{({\rm as})}_\phi(1,q p^{-\theta})
=q^{-4+\eta_{L4}}\,\tilde{G}^{({\rm as})}_\phi(p q^{-1/\theta},1)\,.
\end{equation}
Only two exponents are independent here since the usual scaling relation
\begin{equation}
  \label{eq:thetascrel}
  \theta=\frac{2-\eta_{L2}}{4-\eta_{L4}}
\end{equation}
must hold for the anisotropy index $\theta$ by consistency.

In the limit $n\to\infty$ with $n\lambda=\textrm{fixed}$,
$\tilde{G}^{({\rm as})}_\phi(p,q)$ reduces to the Gaussian propagator
$ \tilde{G}^{(0)}(p,q)$ pertaining to the
Hamiltonian~(\ref{eq:EH}) with $\lambda=\tau_{\rm LP}=\rho_{\rm
  LP}=0$. We have
\begin{equation}\label{eq:Sphl}
\lim_{\stackrel{n\to \infty}{n\lambda={\rm const}}}\tilde{G}^{({\rm
    as})}_\phi(p,q)=\tilde{G}^{(0)}(p,q)\equiv \frac{1}{p^2+q^4}\,.
\end{equation}

At order $1/n$, self-consistent equations must be solved which were
discussed in I and need not be repeated here. To this end, we looked
for solutions of the scaling form~(\ref{eq:Gold}), utilizing the
ansatzes
\begin{equation}\label{eq:Mainr}\fl
\eta_{L2}=\frac{\eta_{L2}^{(1)}}{n}+\Or\big(n^{-2}\big),\quad
\eta_{L4}=\frac{\eta_{L4}^{(1)}}{n}+\Or\big(n^{-2}\big),
\quad\quad
\theta=\frac{1}{2}+{\theta^{(1)}\over n}+\Or\big(n^{-2}\big),
\end{equation}
together with corresponding $1/n$ expansions for the scaling functions in
equation~(\ref{eq:Gold}) and the relation
\begin{equation}
  \label{eq:thetaetas}
  \theta^{(1)}=\frac{\eta_{L4}^{(1)}}{8}-\frac{\eta_{L2}^{(1)}}{4}
\end{equation}
implied by the scaling law~(\ref{eq:Gold}). This led to consistency
conditions (equations (27) and (28) of I) from which we obtained the results
\begin{equation}\label{eq:E2c}
\eta_{L2}^{(1)}=\frac{K_{d-m}}{d-m}
\int_{\bm{q}}^{(m)}\frac{2{\mathcal P}_1(q^4)}{(1+q^4)^3}\;\frac{1}{I(1,q)}
\end{equation}
and
\begin{equation}\label{eq:E41}
\eta_{L4}^{(1)}=\frac{K_m}{4m(m+2)}\,
\int_{\bm{p}}^{(d-m)}\frac{8{\mathcal P}_2(p^2)}{(p^2+1)^5}\,
\frac{1}{ I(p,1)}\;.
\end{equation}
Here $K_{d-m}$ and $K_m$, defined by
\begin{equation}
  \label{eq:KD}
  \quad K_D\equiv\frac{S_D}{(2\pi)^D}\quad\textrm{with}\quad S_D=
  \frac{2\pi^{D/2}}{\Gamma(D/2)}\;,
\end{equation}
where $S_D$ is
the area of a unit sphere in $D$ dimensions, are conventional factors
resulting from the angular integrations at $D=d-m$ and $m$. Further, ${\mathcal
  P}_1(q^4)$ and ${\mathcal P}_2(p^2)$ denote the polynomials
\begin{equation}\label{eq:Pcal1}
{\mathcal P}_1(q^4)=4-(d-m)(1+q^4)
\end{equation}
and
\begin{eqnarray}\label{eq:Pcal2}
{\mathcal P}_2(p^2)&=&3(8-m)(6-m)+5(m^2+2m-96)p^2\nonumber\\ &&\strut
+(m^2+50m+144)p^4-m(m+2)p^6\;.
\end{eqnarray}
Finally, $I(p,q)$ represents the analogue of Ma's ``elementary bubble''
\raisebox{-3pt}{\includegraphics[width=40pt]{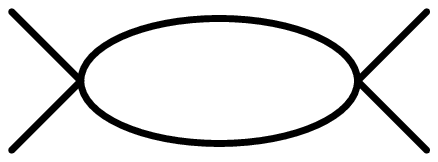}}\; \cite{Ma73} for the
LP:
\begin{equation}\label{eq:Ipq}
I(p,q)=\int_{\bm{p}'}^{(d-m)}\int_{\bm{q}'}^{(m)}
\frac{1}{{p^\prime}^2+{q^\prime}^4}\;
\frac{1}{|\bm{p}'+\bm{p}|^2+|\bm{q}'+\bm{q}|^4}\;,
\end{equation}
whose homogeneity property
\begin{equation}
  \label{eq:Scali}
I(p,q)=p^{-\epsilon}\,I\big(1,qp^{-1/2}\big)=q^{-2\epsilon}
\,I\big(pq^{-2},1\big)\;,\quad \epsilon=4-d+m/2\;,
\end{equation}
we recall for later use.

Let us first show that the results (\ref{eq:E2c}) and (\ref{eq:E41})
for the $\Or(1/n)$ coefficients can be rewritten as
\begin{equation}\label{eq:E2n}
\eta_{L2}^{(1)}=\frac{K_m}{2(d-m)}
\int_{\bm{p}}^{(d-m)}\frac{1}{p^2+1}\;\nabla_{\bm p}^2\,\frac{1}{I(p,1)}
\end{equation}
and
\begin{equation}\label{eq:E4n}
\eta_{L4}^{(1)}=\frac{K_{d-m}}{2m(m+2)}\,
\int_{\bm{q}}^{(m)}\frac{1}{1+q^4}\;\nabla_{\bm q}^4\,\frac{1}{I(1,q)}\;,
\end{equation}
respectively, where $\nabla_{\bm q}^4\equiv\big(\nabla_{\bm
  q}^2\big)^2$. These representations of the coefficients
$\eta_{L2}^{(1)}$ and $\eta_{L4}^{(1)}$ are well suited for determining
the $\Or(\epsilon^2/n)$ contributions to the exponents $\eta_{L2}$ and
$\eta_{L4}$. They will be employed as starting point in our proof of
consistency with the $\epsilon$-expansion results of \cite{DS00a,SD01}
given in the next section.

To derive these representations, note first that the actions of the
$D$-dimensional Laplacian
$\nabla_{\bm{K}}^2=\sum_{\gamma=1}^D\partial^2/\partial K_\gamma^2$ and its square
$\nabla_{\bm{K}}^4$ on functions $f(K^2)$ of $K^2=\sum_{\gamma=1}^DK_\gamma^2$
can be written as
\begin{equation}
\nabla_{\bm K}^2 f(K^2)=2Df'(K^2)+4K^2f''(K^2)
\end{equation}
and
\begin{equation}\fl
\nabla_{\bm{K}}^4f(K^2)=
4D(D+2)f''(K^2)+16(D+2)K^2f^{(3)}(K^2)+16K^4f^{(4)}(K^2)\,,
\end{equation}
where $f^{(s)}(.)$ means the $s$th derivative of the function  $f(.)$.

Using these relations, it is straightforward to
see that the rational functions appearing in the integrands of
(\ref{eq:E2c}) and (\ref{eq:E41}) can be expressed as
 \begin{equation}\label{eq:n4}
 \frac{2\,\mathcal{P}_1(q^4)}{(1+q^4)^3}=
 \nabla _{\bm P}^2 \left.\frac{1}{P^2+q^4}\right|_{P^2=1}
\end{equation}
and
\begin{equation}\label{eq:N4}
 \frac{8\,\mathcal{P}_2(p^2)}{ (1+p^2)^5}=
 \nabla _{\bm Q}^4 \left.\frac{1}{p^2+Q^4}\right|_{Q^2=1}\,.
\end{equation}

We now insert these results into equations~(\ref{eq:E2c}) and
(\ref{eq:E41}), use hyper-spherical coordinates for the integrals
$\int_{\bm{q}}^{(m)}$ and $\int_{\bm{p}}^{(d-m)}$, make the changes of
variables $q\to p=q^{-2}$ and $p\to q=p^{-1/2}$ in the radial
integrals over $q$ and $p$, and utilize the scaling
property~(\ref{eq:Scali}) to express $I(1,p^{-1/2})$ and $I(q^{-2},1)$
in terms of $I(p,1)$ and $I(1,q)$ respectively. The derivative term on
the right-hand side of equation~(\ref{eq:n4}) becomes
$p^2\nabla_{\bm{P}}^2\,(P^2p^2+1)^{-1}|_{P=1}
=p^2\,\nabla_{\bm{p}}^2\,(p^2+1)^{-1}$. The $\nabla_{\bm{Q}}^4$ term
in equation~(\ref{eq:N4}) transforms in a corresponding fashion. One
thus arrives at expressions that agree with equations~(\ref{eq:E2n})
and (\ref{eq:E4n}) except that the derivatives act to the left.
Integration by parts then yields the claimed results.

A straightforward, though important first application of them is to
show that the $\epsilon$~expansion of the coefficients
$\eta_{L2}^{(1)}$ and $\eta_{L4}^{(1)}$ starts at order $\epsilon^2$:
\begin{equation}
  \label{eq:etaL24eps}
  \eta_{L2,4}^{(1)}=\eta_{L2,4}^{(1,2)}\,\epsilon^2+\Or(\epsilon^3)\;.
\end{equation}
To see this, note that $I(p,q)$ has a Laurent expansion about
$\epsilon=0$ of the form
\begin{equation}
  \label{eq:Ipqexp}
  I(p,q)=\frac{I_{-1}}{\epsilon}+I_0(p,q)+\Or(\epsilon)
\end{equation}
with a momentum-independent residuum $I_{-1}$, given by
\begin{equation}
  \label{eq:Imin1}
  I_{-1}=(4\pi)^{-(8+m)/4}\,\frac{\Gamma(m/4)}{\Gamma(m/2)}
\end{equation}
according to references \cite{DS00a,SD01} (see equations (7), (24) and
(89) of \cite{DS00a} or (38) and (39) of \cite{SD01}, where $I_{-1}$
was denoted $F_{m,0}$). Hence
\begin{equation}
  \label{eq:deriviinveps}
  \nabla_{\bm{p}}^2\,\frac{1}{I(p,q)}=-\epsilon^2\,
  \frac{\nabla_{\bm{p}}^2\,I_0(p,q)}{I_{-1}^2}  +\Or(\epsilon^3)
\end{equation}
with $\nabla_{\bm{p}}^2\,I_0(p,q)
=[\nabla_{\bm{p}}^2\,I(p,q)]_{\epsilon=0}$. Analogous results with
$\nabla_{\bm{p}}^2$ replaced by $\nabla_{\bm{q}}^4$ hold. Thus both
$\eta_{L2}^{(1)}$ and $\eta_{L4}^{(1)}$ are indeed of order
$\epsilon^2$, and for their $\Or(\epsilon^2)$ expansion coefficients
$\eta^{(1,2)}_{L2,4}$ we obtain from equations (\ref{eq:E2n}),
(\ref{eq:E4n}) and (\ref{eq:deriviinveps}) the results
\begin{equation}
  \label{eq:etaL212}
  \eta^{(1,2)}_{L2}=-\frac{K_m}{8-m}\, \frac{1}{I_{-1}^2}
  \int_{\bm{p}}^{(4-m/2)}
  \frac{1}{p^2+1}\,\nabla_{\bm{p}}^2\,I(p,1)\big|_{\epsilon=0}
\end{equation}
and
\begin{equation}
  \label{eq:etaL412}
  \eta^{(1,2)}_{L4}=-\frac{K_{4-m/2}}{2m(m+2)}\, \frac{1}{I_{-1}^2}
  \int_{\bm{q}}^{(m)}
  \frac{1}{1+q^4}\,\nabla_{\bm{q}}^4\,I(1,q)\big|_{\epsilon=0} \;.
\end{equation}

\section{Epsilon expansions of $\eta_{L2}^{(1)}$ and
  $\eta_{L4}^{(1)}$}\label{sec:proof}

We are now ready to present the announced proof of consistency. We
shall show that the $\epsilon$ expansions of the correlation exponents
$\eta_{L2}$ and $\eta_{L4}$ are related to the
coefficients $\eta_{L2}^{(1)}$ and $\eta_{L4}^{(1)}$ via
\begin{eqnarray}\label{eq:etaconsist}
  \label{eq:consistclaim}
  \eta_{L2,4}=\frac{n+2}{(n+8)^2}\,\eta_{L2,4}^{(1)}+\Or(\epsilon^3)\;.
\end{eqnarray}
Since the prefactor $(n+2)/(n+8)^{2}$ on the right-hand side reduces
to $1/n$ in the large-$n$ limit, consistency between
the $\epsilon$ expansions to second order and the $1/n$ expansions to
first order is an immediate consequence.

The $\Or(\epsilon^2)$ results of references \cite{DS00a,SD01} for the
exponents $\eta_{L2,4}$ involved single integrals, which for general
$m$ had to be computed by numerical means. In the notation of the
second of these publications, they read
\begin{equation}\label{eq:jphidef}
j_\phi(m)\equiv B_m
\,{\int_0^\infty}{\rmd}\upsilon\,\upsilon^{m-1}\,
\Phi^3(\upsilon;m,d^*)\;,
\end{equation}
and
\begin{equation}\label{eq:jsigmadef}
  j_\sigma(m)\equiv
  B_m\,{\int_0^\infty}{\rmd}\upsilon\,\upsilon^{m+3}\,
  \Phi^3(\upsilon;m,d^*)\;,
\end{equation}
where
\begin{equation}\label{eq:Bm}
B_m =\frac{S_{4-m/2}\,S_m}{I_{-1}^2}
\end{equation}
and
\begin{equation}
  \label{eq:Phi}
  \Phi(\upsilon;m,d)\equiv G^{(0)}(1,\upsilon)\;.
\end{equation}

The latter is the scaling function associated with the free
position-space propagator $G^{(0)}(r,z)$, whose scaling properties
\begin{equation}
  \label{eq:Gfreehomogene}
  G^{(0)}(r,z)=r^{-2+\epsilon}\,G^{(0)}\big(1,zr^{-1/2}\big)
  =z^{-4+2\epsilon}\,G^{(0)}\big(rz^{-2},1\big)
\end{equation}
we recall. For general values of $m$ and $d$, it is a difference of
two generalized hypergeometric functions $ _1F_2$. This is why no analytic
results for the integrals $j_\phi(m)$ and $j_\sigma(m)$ are available for
general $m$. In reference~\cite{SD01}, the reader may find numerical
results for them at $m=1,2,\ldots,7$ along with analytical ones for
$m=2$ and $m=6$.

To prove the relations~(\ref{eq:etaconsist}), we must show that
\begin{equation}
  \label{eq:etaL2eps}
  \eta^{(1,2)}_{L2}=\frac{2}{8-m}\,j_\phi(m)
\end{equation}
and
\begin{equation}
  \label{eq:etaL4eps}
\eta^{(1,2)}_{L4}=-\frac{1}{2m(m+2)}\,j_\sigma(m)\;.
\end{equation}

Let us start from equation~(\ref{eq:etaL212}).
Its integral $\int_{\bm{p}}^{(4-m/2)}$ has the form of a
scalar product $\langle f|g\rangle$ in $L_2(\mathbb{R}^{4-m/2})$, the
space of square integrable functions, that is evaluated in the
$\bm{p}$-representation.  In $\bm{r}$-space the bra $\langle f|$ is
represented by $f(r)^*\equiv \langle f|\bm{r}\rangle$, where $f(r)$ is
the Fourier $\bm{q}$-transform of $G^{(0)}(r,z)$, taken at an arbitrary unit
$\bm{q}$-vector $\hat{\bm{q}}$. Performing the angular integrations in
the required $m$-dimensional integral is straightforward and yields
\begin{equation}
  \label{eq:fgrspace}
 f(r)=\int d^mz\, G^{(0)}(r,z)\,\e^{\rmi\hat{\bm{q}}\cdot\bm{z}}= (2\pi)^{m/2}
 \int_0^\infty dz\,z^{m/2}\,J_{\frac{m}{2}-1}(z)\,G^{(0)}(r,z)\;,
\end{equation}
where from now on $\epsilon$ is set to zero in $G^{(0)}(r,z)$.

Likewise, $\langle\bm{r}|g\rangle\equiv g(r)$ is the Fourier
$\bm{q}$-transform of the function
$(-r^2)\,[G^{(0)}(r,z)]^2$ for $\bm{q}=\hat{\bm{q}}$. In the resulting
expression for
\begin{eqnarray}
  \label{eq:scprodfg}
  \langle f|g\rangle&=&-(2\pi)^m\int d^{4-m/2}r\,\int_0^\infty
  dz\int_0^\infty dz'\,r^2\,(zz')^{m/2}\, J_{\frac{m}{2}-1}(z)\,G^{(0)}(r,z)
  \nonumber \\ &&\strut\times J_{\frac{m}{2}-1}(z')\,[G^{(0)}(r,z')]^2\,
\end{eqnarray} substitute the first of the
scaling forms~(\ref{eq:Gfreehomogene}) along with
equation~(\ref{eq:Phi}). We then make the changes of
variables $r\to \upsilon \equiv zr^{-1/2}$ in the radial part of the
integration over $\bm{r}$ and  $z'\to
\zeta=z'/z$. The resulting integral over $z$ is the special case  of
the  closure relation for Bessel functions\footnote{See, for instance,
  \cite{Hugh95}}
\begin{equation}
  \label{eq:Besseljorth}
  \int_0^\infty dz\,z\,J_\mu(\zeta z)\, J_\mu(b z) = \delta(\zeta-b)/\zeta
\end{equation}
with $\mu=m/2-1$ and $b=1$. The integral over  $\zeta$ can now be
performed. Substituting the result into equation~(\ref{eq:etaL212}) and
noting (\ref{eq:jphidef}) and (\ref{eq:Bm}) then gives the asserted
result~(\ref{eq:etaL2eps}) for the coefficient $\eta^{(1,2)}_{L2}$.

The corresponding expression~(\ref{eq:etaL4eps}) for
$\eta^{(1,2)}_{L4}$ can be proven in an analogous fashion. The
integral $\int_{\bm{q}}^{(m)}$ is a scalar product $\langle
h|w\rangle$ in $L_2(\mathbb{R}^m)$ between $h(z)$, the Fourier
$\bm{p}$-transform of $G(r,z)$, and $w(z)$, that of $z^4\,[G^{(0)}(r,z)]^2$,
taken at a unit $\bm{p}$-vector $\hat{\bm{p}}$. We perform the angular
integrals in the Fourier integrals $\int d^{4-m/2}r$ and $\int
d^{4-m/2}r'$, and make the changes of variables $z\to\upsilon
=zr^{-1/2}$ and $r'\to \zeta=r'/ r$. The integral over $r$ is of the
form~(\ref{eq:Besseljorth}) with $\mu=1-m/4$ and $b=1$. Once the
integral over $\zeta$ is performed, the desired result follows from
equations~(\ref{eq:etaL412}), (\ref{eq:Bm}), and (\ref{eq:jsigmadef}).

\section{Concluding remarks} \label{sec:concl}

In this paper we have shown that the large-$n$ expansion yields  $\Or(1/n)$
results for the correlation exponents $\eta_{L2}$ and
$\eta_{L4}$ and the related anisotropy exponent $\theta$ for general
$m$, which are fully consistent with the $\epsilon$-expansion of
references~\cite{DS00a,SD01}. In view of the long-standing
discrepancies mentioned in the Introduction and the great technical
challenges encountered in both expansion methods beyond lowest order,
the established consistency is very gratifying, providing nontrivial
checks of the results of both expansions given in
references~{\cite{DS00a,SD01} and \cite{SPD05}, respectively.

\ack

YuMP was supported in part by Russian Foundation of Basic
Research (Grant 07-01-00692-a).

\end{document}